\documentclass{svmult}

\usepackage{helvet}         % selects Helvetica as sans-serif font
\usepackage{courier}        % selects Courier as typewriter font
\usepackage{type1cm}        % activate if the above 3 fonts are
\usepackage{makeidx}         % allows index generation
\usepackage{graphicx}        % standard LaTeX graphics tool
\usepackage{multicol}        % used for the two-column index
\usepackage[bottom]{footmisc}% places footnotes at page bottom
\usepackage{mathtools}
\usepackage{amssymb}
\usepackage{latexsym}

\makeindex             % used for the subject index
\begin{document}

\title*{Block local elimination algorithms for solving sparse discrete
optimization problems}

\author{Alexander Sviridenko and Oleg Shcherbina}

\institute{A. Sviridenko \at
  Faculty of Mathematics and Computer Science \\
  Tavrian National University, Vernadsky Av. 4,  Simferopol 95007  Ukraine \\
  \email{oleks.sviridenko@gmail.com}
  \and
  O. Shcherbina \at
  Faculty of Mathematics, University of Vienna \\
  Nordbergstrasse 15,  A-1090 Vienna,  Austria \\
  \email{oleg.shcherbina@univie.ac.at}
}

\maketitle

\abstract{Block elimination algorithms for solving sparse discrete
  optimization problems are considered. The numerical example is
  provided. The benchmarking is done in order to define real
  computational capabilities of block elimination algorithms combined
  with SYMPHONY solver. Analysis of the results show that for
  sufficiently large number of blocks and small enough size of
  separators between the blocks for staircase integer linear
  programming problem the local elimination algorithms in combination
  with a solver for solving subproblems in blocks allow to solve such
  problems much faster than used solver itself for solving the whole
  problem. Also the capabilities of postoptimal analysis (warm
  starting) are considered for solving packages of integer linear
  programming problems for corresponding blocks.}

\section{Introduction}

The use of discrete optimization (DO) models and algorithms makes
it possible to solve many practical problems, since the discrete
optimization models correctly represent the nonlinear dependence,
indivisibility of an objects, consider the limitations of logical type
and all sorts of technology requirements, including those that have
qualitative character. But unfortunately, most of the interesting
problems are in the complexity class $NP$-hard and may require
searching a tree of exponential size in the worst case. Many real DOPs
from OR applications contain a huge number of variables and/or
constraints that make the models intractable for currently available
solvers. Usually, DOPs from applications have a special structure, and
the matrices of constraints for large-scale problems have a lot of
zero elements (sparse matrices), and the nonzero elements of the
matrix often fall into a limited number of blocks. The block form of
many DO problems is usually caused by the weak connectedness of
subsystems of real-world systems.

\begin{figure}[htbp]
\centering
\includegraphics[scale=0.5]{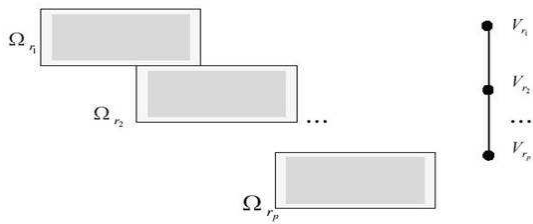}
\caption{Quasi-block structure.}\label{QuasiBlockStructure}
\end{figure}

Among the block structures let us pay particular attention to the
block-tree structure, a special case of which is a staircase or
quasiblock structure (Fig. \ref{QuasiBlockStructure}). The problems of
optimal reservation of hotel rooms \cite{Soa83} have the quasiblock structure,
similar to the previous \textit{temporal knapsack problems} \cite{Bartlett}, recently
received the application in solving problems of the prior reservation
of computing resources in the \textit{Grid Computing}.

One of the promising ways to exploit sparsity in the constraint matrix
of DO problems are local elimination algorithms (LEA)\cite{Shc2}, including
local decomposition algorithms \cite{Soa83}, nonserial dynamic programming
(NSDP) \cite{Bertele} algorithms, \cite{Soa07}. To extract special block structures there
are such promising graph-based decomposition approaches as methods of
tree decomposition \cite{SoaKibern07}. The purpose of this paper to define real
computational capabilities of block elimination algorithms combined
with modern solvers.

\section{Local elimination algorithms for solving discrete problems}

\subsection{General scheme of local elimination algorithms}

In the papers \cite{Shc2}, \cite{Soa09} considered the general class of \textit{local elimination
algorithms for computing information}, that have decomposition approach and that allow to
calculate some \textit{global} information about a solution of the entire
problem using \textit{local} computations.

A local elimination algorithm (LEA) eliminates local elements of the
problem’s structure defined by the structural graph by computing and
storing local information about these elements in the form of new
dependencies added to the problem.

The local elimination procedure consists of two parts:
\begin{itemize}
  \item [1.] The \textit{forward part} eliminates elements, computes and stores
    local solutions, and finally computes the value of the objective
    function;
  \item [2.] The \textit{backward part} finds the global solution of the whole
    problem using the tables of local solutions; the global solution
    gives the optimal value of the objective function found while
    performing the forward part of the procedure.
\end{itemize}

The algorithmic scheme of the LEA is a directed acyclic graph (DAG) in
which the vertices correspond to the local subproblems and the edges
reflect the informational dependence of the subproblems on each other.

It is important that aforementioned methods use just the \textit{local
  information} (i.e., information about elements of given element’s
neighborhood) in a process of solving discrete problems. Thus local
elimination algorithms allow to calculate some global information
about a solution of the entire problem using local computations.

The structure of discrete optimization problems is determined either
by the original elements (e.g., variables) with a system of
neighborhoods specified for them with help of structural graph and
with the order of searching through those elements using a LEA or by
various derived structures (e.g., block or tree-block structures).

\subsection{Local elimination algorithms}

Consider LEA in details for solving sparse problems of integer linear
programming in the case when structural graph is \textit{interaction
  graph} of variables, which is also called \textit{constraint graph}.

Consider the integer linear programming (ILP) problem with binary variables

\begin{equation}
f(X) = CX = \sum_{j=1}^{n} c_jx_j \rightarrow \max
\end{equation}

subject to constraints

\begin{equation}
\sum_{j=1}^na_{ij}x_j \leq b_i, i=1,2,\dots,m,
\end{equation}

\begin{equation}
  x_j = 0,1, j=1,2,\dots,n.
\end{equation}

\begin{definition} Variables $x$ and $y$ interact in ILP problem with
  constraints if they both appear in the same constraint.
\end{definition}

\begin{definition} \textit{Interaction graph} of the ILP problem is an undirected graph
  $G=(X, E)$, such that
  \begin{itemize}
  \item[1.] Vertices $X$ of $G$ correspond to variables of the ILP problem;
  \item[2.] Two vertices of $G$ are adjacent iff corresponding
    variables interact.
  \end{itemize}
\end{definition}

Further, we shall use the notion of vertices that correspond
one-to-one to variables.

\begin{example} Consider an ILP problem with binary
variables:
\begin{equation}
  \begin{aligned}
  &2x_1+3x_2+~x_3+5x_4+4x_5+6x_6+~x_7&\rightarrow \max\\
  \end{aligned}
\end{equation}
subject to constraints
\begin{equation}
  \begin{aligned}
  &3x_1+4x_2+~~x_3~~~~~~~~~~~~~~~~~~~~~~~~~~&~~~~~\leq 6,
  \end{aligned}
\end{equation}
\begin{equation}
  \begin{aligned}
  &~~~~~~~~2x_2+3x_3+3x_4~~~~~~~~~~~~~~~~~~&~~~~~\leq 5,
  \end{aligned}
\end{equation}
\begin{equation}
  \begin{aligned}
  &~~~~~~~~2x_2~~~~~~~~~~~~~~~~+3x_5~~~~~~~&~~~~~~~~\leq 4,
  \end{aligned}
\end{equation}
\begin{equation}
  \begin{aligned}
  &~~~~~~~~~~~~~~~~2x_3~~~~~~~~~~~~~~~~+3x_6+2x_7&\leq 5,
  \end{aligned}
\end{equation}
\begin{equation}
  \begin{aligned}
  &~x_j=0,1,~j=1,\ldots, 7.~~~~~~~~~~~~~~~~~~~~~~~~~~~~~~~~~
  \end{aligned}
\end{equation}
\end{example}

\begin{definition} Two vertices $x$ and $y$ of $G=(X,E)$ are called
  \textit{neighbors} if $(x,y) \in E$.
\end{definition}

\begin{definition} Set of variables interacting with a variable $x \in X$
  is denoted by $Nb(x)$ and called the \textit{neighborhood} of the
  variable $x$. For corresponding vertices a neighborhood of a vertex
  $x$ is a set of vertices of interaction graph $G=(X,E)$ that are
  linked by edges with $x$. Denote the latter neighborhood as
  $Nb_G(x)$.
\end{definition}

The solution of a sparse discrete optimization problem (1) -- (3)
whose structure is described by an undirected interaction graph $G =
(X,E)$ with help of LEA was described in \cite{SavSoa} in details. Given an
ordering $x_1, x_2, \dots, x_n$, the LEA proceeds in the following
way: it subsequently eliminates $x_1, x_2, \dots, x_n$ in the current
graph and computes an associated local information about vertices from
$h_i(Nb(x_i))$. This process creates a sequence of elimination graphs:
$G^0=G, G^1,\dots,G^j,\dots,G^n$, where $G^n=\emptyset$.

The process of interaction graph transformation corresponding to the
LEA scheme is known as \textit{elimination game} which was first introduced
by Parter as a graph analogy of Gaussian elimination.

\subsection{Block local elimination algorithms}

The local elimination procedure can be applied to elimination of not
only separate variables but also to sets of variables and can use the
so called elimination of variables in blocks, which allows to
eliminate several variables in block.

Applying the method of merging variables into meta-variables allows to
obtain \textit{condensed} or meta-DOPs which have a simpler structure. If the
resulting meta-DOP has a nice structure (e.g., a tree structure) then
it can be solved efficiently.

An \textit{ordered partition} of a set $X$ is a decomposition of $X$
into ordered sequence of pairwise disjoint nonempty subsets whose
union is all of $X$.

In general, graph partitioning is $NP$-hard. Since graph partitioning
is difficult in general, there is a need for approximation
algorithms. A popular algorithm in this respect is
\textbf{MeTiS}\footnote{http://www-users.cs.umn.edu/~karypis/metis}, which has
a good implementation available in the public domain.

An important special case of partitions are so-called
\textit{blocks}. Two variables are \textit{indistinguishable} if they
have the same closed neighborhood. A \textit{block} is a maximal set
of indistinguishable vertices. The blocks of $G$ partition $X$ since
indistinguishability is an \textit{equivalence relation} defined on
the original vertices. The corresponding graph is called
\textit{condensed graph}, which is a merged form of original
graph. Not formally, a condensed graph is formed by merging all
vertices with the same neighborhoods into a single meta-node
(supervariable).

An equivalence relation on a set induces a partition on it, and also
any partition induces an equivalence relation. Given a graph
$G=(X,E)$, let $X$ be a partition on the vertex set $X$: $X = {x_1,
  x_2, \dots, x_p}$, $p \leq n$, where $x_l=X_{K_l}$($K_l$ is a set of
indices corresponding to $x_l$, $l=1,\dots,p$). For this ordered
partition $X$, the DOP (1) -- (3) can be solved by the LEA using
\textit{quotient interaction graph} $G$.

That is, $\cup_{i=1}^{p} \mathbf{x_i} = X$ and $\mathbf{x_i} \cap
\mathbf{x_k} = \emptyset$ for $i \ne k$. We define the
\textit{quotient graph} of $G$ with respect to the partition
$\mathbf{X}$ to be the graph

\[
\mathbf{G} = G / \mathbf{X}  = (\mathbf{X}, \mathcal{E}),
\]

where $(\mathbf{x_i},~\mathbf{x_k}) \in \mathcal{E}$ if and only if
$Nb_G(\mathbf{x_i}) \cap \mathbf{x_k} \ne \emptyset$.

The quotient graph $\mathbf{G}(\mathbf{X},\mathcal{E})$ is an
equivalent representation of the interaction graph $G(X,E)$, where
$\mathbf{X}$ is a set of blocks (or indistinguishable sets of
vertices), and $\mathcal{E} \subseteq \mathbf{X} \times \mathbf{X}$ be
the edges defined on $\mathbf{X}$. A \textit{local block elimination}
scheme is one in which the vertices of each block are eliminated
contiguously. As an application of a clustering technique we consider
below a block local elimination procedure where the elimination of the
block (i.e., a subset of variables) can be seen as the merging of its
variables into a meta-variable.

\textbf{A. Forward part}

Consider first the block $\mathbf{x_1}$. Then
\[
\max_{X} \{C_N X_N|A_{iS_i}X_{S_i} \leq b_i,~  i \in  M,~
x_{j}=0,1,~ j \in N\} =
\]
\[
\max_{X_{K_2},\ldots,X_{K_p}} \{C_{N-K_1} X_{N-K_1}+
h_1(Nb(X_{K_1})|A_{iS_i}X_{S_i} \leq b_i,~ i \in M-U_1,\]
\[ x_{j}=0,1,~j \in N-K_1\}
\]
 where $U_1=\{i:S_i \cap K_1 \ne \emptyset\}$ and
\[
h_1(Nb(X_{K_1})) = \max_{X_{K_1}} \{C_{K_1}
X_{K_1}|A_{iS_i}X_{S_i} \leq b_i,~  i \in U_1,~ x_{j}=0,1,~ x_j \in Nb[\mathbf{x_1}]\}.
\]
The first step of the local block  elimination procedure consists of
solving, using complete enumeration of $X_{K_1}$, the following
optimization problem
\begin{equation}\label{h_1}
h_1(Nb(X_{K_1})) = \max_{X_{K_1}} \{C_{K_1} X_{K_1}|A_{iS_i}X_{S_i}
\leq b_i,~  i \in U_1,~ x_{j}=0,1,~ x_j \in Nb[\mathbf{x_1}]\},
\end{equation}
and storing the optimal local solutions $X_{K_1}$ as a function of
the neighborhood $of X_{K_1}$, i.e., $X_{K_1}^*(Nb(X_{K_1}))$.

The maximization of $f(X)$ over all feasible assignments
$Nb(X_{K_1})$, is called the \textit{elimination of the block} (or
meta-variable) $X_{K_1}$. The optimization problem left after the
elimination of $X_{K_1}$ is:
\[
\max_{X-X_{K_1}} \{C_{N-K_1} X_{N-K_1}+
h_1(Nb(X_{K_1}))|A_{iS_i}X_{S_i} \leq b_i,~ i \in M-U_1, x_{j}=0,1,~ j \in N-K_1 \}.
\]

Note that it has the same form as the original problem, and the
tabular function $h_1(Nb(X_{K_1}))$ may be considered as a new
component of the modified objective function. Subsequently, the same
procedure may be applied to the elimination of the blocks --
meta-variables $\mathbf{x_2}=X_{K_2},\ldots,\mathbf{x_p}=X_{K_p}$, in
turn. At each step $j$ the new component $h_{\mathbf{x_j}}$ and
optimal local solutions $X^{*}_{K_j}$ are stored as functions of
$Nb(X_{K_j} \mid X_{K_1}, \ldots, X_{K_{j-1}})$, i.e., the set of
variables interacting with at least one variable of $X_{K_j}$ in the
current problem, obtained from the original problem by the elimination
of $X_{K_1}, \ldots, X_{K_{j-1}}$. Since the set $Nb(X_{K_{p}} \mid
X_{K_{1}}, \ldots, X_{K_{p-1}})$ is empty, the elimination of
$X_{K_{p}}$ yields the optimal value of objective $f(X)$.

\textbf{B. Backward part.}

This part of the procedure consists of the consecutive choice of
$X^{*}_{K_p}$, $X^{*}_{K_{p-1}},\ldots, X^{*}_{K_{1}}$, i.e., the
optimal local solutions from the stored tables $ X^{*}_{K_{1}}
(Nb(X_{K_{1}})), X^{*}_{K_{2}}(Nb(X_{K_{2}} \mid X_{K_{1}})), \ldots,
X^{*}_{K_{p}} \mid X_{K_{p-1}},\ldots, X_{K_{1}}$.

Underlying DAG of the local block elimination procedure contains nodes
corresponding to computing of functions
$h_{\mathbf{x_i}}(Nb_{G_{\mathbf{X}}^{(i-1)}}(\mathbf{x_i}))$ and is a
\textit{generalized elimination tree}.

\begin{example} Consider a DO problem (4) -- (9) from example 1
and an ordered partition of the variables of the set $X$ into blocks:
$\mathbf{x_1} = \{x_5\}, \mathbf{x_2} = \{x_1, x_2, x_4\},
 \mathbf{x_3} = \{x_6, x_7\}, \mathbf{x_4} = \{x_3\}$. For the ordered
partition $\{x_1, x_2, x_3, x_4\}$, this
DO problem may be solved by the LEA. Initial interaction
graph with partition presented by dashed lines is shown in Fig. \ref{BlockGraph} (a),
quotient interaction graph is in Fig. \ref{BlockGraph} (b), and the DAG of the block
local elimination computational procedure is shown in Fig. \ref{DAG_BlockGraph}.

\begin{figure}[htbp]
\centering
\includegraphics[scale=0.7]{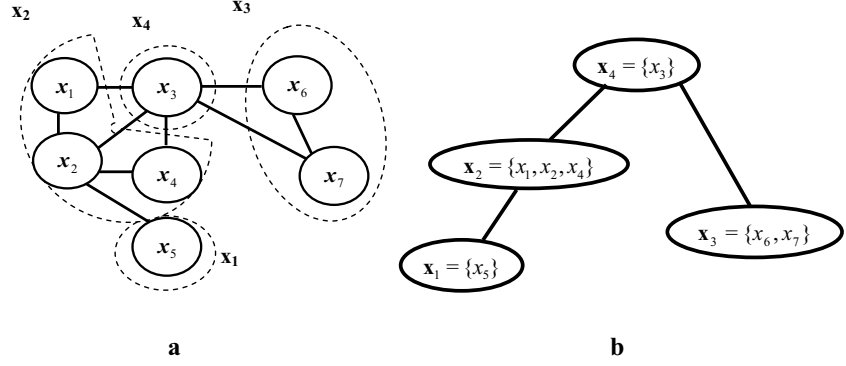}
\caption{Interaction graph of the DOP with partition (dashed) (a) and
  quotient interaction graph (b) (example 2).}\label{BlockGraph}
\end{figure}
\begin{figure}[htbp]
\centering
\includegraphics[scale=0.7]{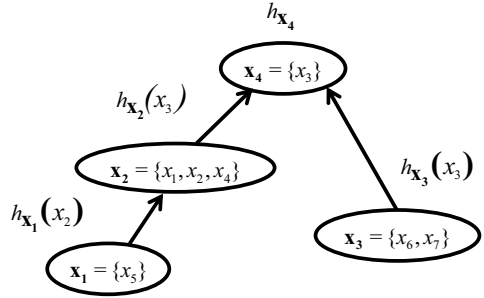}
\caption{The DAG (generalized elimination tree) of the local block
  elimination computational procedure for the DO problem (example
  2).}\label{DAG_BlockGraph}
\end{figure}

\textbf{A. Forward part}

Consider first the block $\mathbf{x_1}=\{x_5\}$. Then $Nb(\mathbf{x_1})=\{x_2\}$.
Solve the following problem containing $x_5$ in the objective and
the constraints:

\[
h_{\mathbf{x_1}}(Nb(\mathbf{x_1}))=\max_{x_5}\{4x_5\mid
2x_2+3x_5 \leq 4, x_j\in \{0,1\}\}
\]

and store the optimal local solutions $\mathbf{x_1}$ as a function of
a neighborhood, i.e., $\mathbf{x_1}^*(Nb(\mathbf{x_1}))$.  Eliminate
the block $\mathbf{x_1}$.  and consider the block
$\mathbf{x_2}=\{x_1,x_2,x_4\}$.  $Nb(\mathbf{x_2})=\{x_3\}$. Now the
problem to be solved is
\[
\begin{aligned}
& h_{\mathbf{x_2}}(x_3)=\max_{x_1,x_2,x_4} \{h_{\mathbf{x_1}}(x_2)+
2x_1+3x_2+5x_4\} \\
\hbox{subject to}\\
&~~~3x_1+4x_2+~x_3~~~~~~~~~~\leq 6,\\
&~~~~~~~~~~~~2x_2+3x_3+3x_4~\leq 5,\\
&~~~~~~~~~~~~~~~~~~~~~~~x_j=0,1,~j=1,2,3,4.~~~~~~~~~~~~~~~~~~~~~~~~~~~~~~~~
\end{aligned}
\]
Build the corresponding table 2.
\begin{center}
\parbox[t]{40mm}{
Table 1. \\{\bf Calculation of $h_{\mathbf{x_1}}(x_2)$}\\[1ex]
\begin{tabular}{|c|c|c|}
\hline
$x_2$&   $h_{\mathbf{x_1}}(x_2)$ & $x_5^*$ \\
\hline
0 &   4 &   1 \\
1 &   0 &   0 \\
\hline
\end{tabular}}
\qquad
\parbox[t]{50mm}{
Table 2. \\{\bf Calculation of $h_{\mathbf{x_2}}(x_3)$}\\[1ex]
\begin{tabular}{|c|c|ccc|}
\hline
 $x_3$& $h_{\mathbf{x_2}}(x_3)$ & $x_1^*$ & $x_2^*$ & $x_4^*$\\
\hline
0 &11&   1 &   0 & 1 \\
1 &6&   1&   0 & 0 \\
\hline
\end{tabular}}
\end{center}
Eliminate the block $\mathbf{x_2}$ and consider the block
$\mathbf{x_3}=\{x_6,x_7\}$.
The neighbor of $\mathbf{x_3}$ is $x_3$:
$Nb(\mathbf{x_3})=\{x_3\}$. Solve the DOP containing $x_3$:

\[
h_{\mathbf{x_3}}(x_3)=\max_{x_6,x_7}\{h_{\mathbf{x_2}}+x_3+6x_6+x_7~|~2x_3+3x_6+2x_7
\leq 5, x_j\in \{0,1\}\}
\]

and build the table 3.
\begin{center}
\parbox[t]{40mm}{
Table 3. \\{\bf Calculation
of $h_{\mathbf{x_3}}(x_3)$}\\[1ex]
\begin{tabular}{|c|c|cc|} \hline
 $x_3$&   $h_{\mathbf{x_3}}(x_3)$ & $x_6^*$ & $x_7^*$\\
\hline
0 &   18 &   1 & 1 \\
1 &   12 &   1 & 0 \\ \hline
\end{tabular}}
\end{center}
Eliminate the block $\mathbf{x_3}$ and consider the block
$\mathbf{x_4}=\{x_3\}$. $Nb(\mathbf{x_4})=\emptyset$. Solve the DOP:
\[
h_{\mathbf{x_4}}=\max_{x_3}\{h_{\mathbf{x_3}}(x_3), x_j\in \{0,1\}\}=18,
\] where $x_3^*=0$.

\textbf{B. Backward part.}

Consecutively find $\mathbf{x_3}^*, \mathbf{x_2}^*, \mathbf{x_1}^*$,
i.e., the optimal local solutions from the stored tables 3, 2, 1.  $
x_3^*=0 \Rightarrow x_6^*=1,~ x_7^*=1$ (table 3); $x_3^*=0 \Rightarrow
x_1^*=1,~ x_2^*=0,~ x_4^*=1$ (table 2); $x_2^*=0 \Rightarrow x_5^*=1$
(table 1).  We found the optimal global solution to be
$(1,~0,~0,~1,~1,~1,~1)$, the maximum objective value is 18.
\end{example}

\section{Research the computational capability of local algorithms}

\subsection{Comparative computational experiment}

Among extremely important research questions about the effectiveness
of local elimination algorithms (LEA), the next one causes special
interest: ``Is the use of LEA in combination with a discrete
optimization (DO) algorithm (for solving problems in the blocks)
consistently more efficient than the standalone use of the DO
algorithm?'' \cite{Soa11}.

Along with the theoretical analysis of performance evaluations, it is
of interest to provide a comparison of LEA combined with other DO
algorithms by using computational experiments. Providing an exhaustive
computational study for all possible combinations of LEA with all
existing DO algorithms (or at least the most efficient ones) is
extremely laborious. This work presents computational
comparisons of the LEA combined with two DO algorithms: a) one of the
least efficient DO algorithms, which is a simple implicit enumeration
algorithm without use of linear relaxation; b) one of the most
effective DO algorithm, as the one adopted by the simplex method in
the unimodular case.

The purpose of these comparisons is to evaluate the behavior of the
LEA in combination with very effective algorithms (``lower bound'')
and weakly effective ones (``upper bound'').

By combining the LEA with implicit enumeration solvers (one of the
least efficient algorithms) \cite{Soa83} we concluded that for
sufficiently large number of blocks and small enough size of
separators between the blocks for staircase ILP
problems the performance is better than the stand alone solver. It
has to be noted, that there are cases that the combination of LEA with
the simplex method for a small number of blocks performs worse than
the simplex method\footnote{The benchmarking was provided with help of
  V.V.Matveev.}. Thus, the practical use of LEA in combination with a
DO solver requires a preliminary machine experiment. Based on this experiment
have to be determined the acceptable parameters for DO problems.

The paper \cite{ilSoa94} presents computational experiment on the
effectiveness (in terms of precision and time to find a solution) of
LEA in combination with approximate algorithms from the package of
applied programs (PAP) ``DISPRO'' developed at the Institute of
Cybernetics of Ukrainian Academy of Sciences.  The studied methods
include lexicographic search, recession vector and random search.  The
object of the experiments was a special class of DO quasi-block
problems, the optimal reservation class, which under certain
conditions are unimodular \cite{Soa83}.  The results showed that if
the efficiency of the PAP ``DISPRO'' algorithms is almost independent
of the matrix condition structure of the DO problem, but is
essentially determined by the dimension of the problem, then the use
of LEA is appropriate for relatively small values of the separators
and sufficient sparsity of the condition matrix.

Additionally, by using the LEA in combination with some DO algorithm
we achieved better accuracy than by using just the DO algorithm. In
some cases the time to solution is several times lower than the
independent calculation with an appropriate DO algorithm.  By
increasing the dimension of the problem and by providing the resource
time restriction the efficiency of the LEA increases.

\subsection{SYMPHONY as a framework for solving mixed integer programming
problems}

The computational capabilities of the LEA in combination with a modern
solver were tested by using
SYMPHONY\footnote{https://projects.coin-or.org/SYMPHONY} as the
implementation framework. SYMPHONY is part of the
COIN-OR\footnote{http://www.coin-or.org} project and it can solve
mixed-integer linear programs (MILP) sequentially or in parallel. We
chose this framework since it is open-source and supports warm
restarts, which implement postoptimal analysis (PA) of ILP problems.

It has to be noted that SYMPHONY does not include an LP-Solver, but
can use, through the Osi interface, third-party solvers such as Clp,
Cplex, Xpress.  Furthermore, SYMPHONY also has a structure-specific
implementations for problems like the traveling salesman problem,
vehicle routing problem, set partitioning problem, mixed postman
problem and others.

\textbf{Warm start technology for implementation of Postoptimal Analysis.}

Warm restarts are used by modern solvers such as Gurobi, SCIP, CBC,
SYMPHONY, and others, in order to implement PA. We use the
capabilities of warm restarts offered by SYMPHONY in order to
reduce the solution time of the subproblems generated by the computational
scheme of LEA \cite{Soa08}.

SYMPHONY implements warm restarting by using a compact description of
the search tree at the time the computation is halted. This
description contains the complete information about the subproblem
corresponding to each node in the search tree, including the branching
decisions that leads to the creation of the node, the list of active
variables and constraints, and warm restart information for the
subproblem itself. All information is compactly stored using
SYMPHONY's native data structures, which store only the differences
between a child and its parent, rather than an explicit description of
every node. In addition to the tree itself, other relevant information
regarding the status of the computation is recorded, such as the
current bounds and best feasible solution.

By using warm restarting, the user can save a warm restart to disk,
read one from the disk, or restart the computation at any point after
modifying parameters or the problem data. Note that the use of the PA
procedure does not always guarantee a positive result.  In the case of
SYMPHONY, it has been observed that, as a rule of thumb, the
warm-restart procedure works best with a slight change in the
conditions of the problem.

\subsection{Benchmarking analysis}

All experimental results were obtained on an Intel Core 2 Duo at 2.66
GHz machine with 2 GB main memory, and running Linux, version
2.6.35-24-generic. SYMPHONY
5.4.1\footnote{http://coin­or.org/download/source/SYMPHONY/} was used
for the LEA implementation. The results are presented in Table
\ref{benchmarking_table}, where $n$ denotes the number of variables,
$m$ the number of constraints, $k$ the number of blocks, $b$ the size
of separator, and underlined is the minimal time of problem solving
for appropriate algorithm. The maximum solving time is denoted by
$TIMEOUT$, and is equal to 2 hours.

\begin{table}[ht!]
  \caption{Run-time (in minutes) of solving generated problems with quasi-block
    structure.}
  \label{benchmarking_table}
  \begin{tabular}{|l|r|r|r|r|r|r|r|}
    \hline\noalign{}
    & \multicolumn{4}{|c|}{Problem parameters} & \multicolumn{3}{|c|}{Solvers}\\
    \hline\noalign{}
    \#  &  $n$  & $m$ &  $k$ & $b$ & SYMPHONY & SYMPHONY + LEA  & SYMPHONY + LEA + PA \\
    \hline\noalign{}
    1 & 180 & 12 & 6 & 1 & 9.503 & 0.028 & \underline{0.026} \\
    2 & 180 & 12 & 6 & 2 & 3.019 & \underline{0.046} & 0.047 \\
    3 & 180 & 12 & 6 & 3 & 1.671 & \underline{0.17} & 0.171 \\
    4 & 180 & 12 & 6 & 4 & 1.164 & 0.493 & \underline{0.485} \\
    5 & 180 & 12 & 6 & 5 & \underline{0.084} & 5.667 & 5.295 \\
    6 & 180 & 50 & 25 & 5 & 1.572 & \underline{0.03} & 0.031 \\
    7 & 180 & 12 & 6 & 6 & \underline{2.7167} & 5.321 & 5.057 \\
    8 & 320 & 16 & 8 & 1 & 9.605 & 0.025 & \underline{0.024} \\
    9 & 320 & 20 & 10 & 1 & 24.435 & 0.029 & \underline{0.026} \\
    10 & 320 & 40 & 20 & 1 & 3.943 & 0.016 & \underline{0.015} \\
    11 & 320 & 20 & 10 & 2 & 18.464 & \underline{0.092} & 0.093 \\
    12 & 320 & 40 & 20 & 2 & 2.519 & \underline{0.048} & 0.049 \\
    13 & 320 & 20 & 10 & 3 & 31 & 0.289 & \underline{0.289} \\
    14 & 320 & 40 & 20 & 3 & 3.382 & 0.158 & \underline{0.157} \\
    15 & 320 & 20 & 10 & 4 & 47.919 & 1.21 & \underline{1.204} \\
    16 & 320 & 20 & 10 & 5 & 14.941 & 3.73 & \underline{3.682} \\
    17 & 320 & 12 & 6 & 6 & 44.556 & \underline{21.534} & 21.636 \\
    18 & 320 & 20 & 10 & 6 & 23.853 & 14.327 & \underline{14.043} \\
    19 & 500 & 50 & 25 & 1 & 90.02 & \underline{0.021} & 0.022 \\
    20 & 500 & 50 & 25 & 2 & TIMEOUT & \underline{0.071} & 0.073 \\
    21 & 500 & 50 & 25 & 3 & TIMEOUT & 0.681 & \underline{0.303} \\
    22 & 500 & 50 & 25 & 4 & TIMEOUT & 0.883 & \underline{0.879} \\
    23 & 500 & 25 & 12 & 5 & TIMEOUT & 7.404 & \underline{7.413} \\
    24 & 500 & 130 & 65 & 5 & TIMEOUT & 0.076 & \underline{0.077} \\
    25 & 500 & 112 & 56 & 6 & TIMEOUT & 0.246 & \underline{0.244} \\
    26 & 800 & 120 & 60 & 1 & TIMEOUT & \underline{0.048} & 0.049 \\
    27 & 800 & 120 & 60 & 2 & TIMEOUT & 0.119 & \underline{0.118} \\
    28 & 800 & 50 & 25 & 3 & TIMEOUT & 0.564 & \underline{0.563} \\
    29 & 800 & 50 & 25 & 4 & TIMEOUT & 2.197 & \underline{2.175} \\
    30 & 800 & 240 & 120 & 4 & TIMEOUT & \underline{0.036} & 0.039 \\
    31 & 800 & 50 & 25 & 5 & TIMEOUT & 8.924 & \underline{8.638} \\
    32 & 800 & 50 & 25 & 6 & TIMEOUT & 31.147 & \underline{30.719} \\
    33 & 800 & 180 & 90 & 6 & TIMEOUT & \underline{0.399} & 0.412 \\
    34 & 1000 & 50 & 25 & 1 & TIMEOUT & \underline{0.073} & 0.075 \\
    35 & 1000 & 50 & 25 & 2 & TIMEOUT & 0.287 & \underline{0.286} \\
    36 & 1000 & 50 & 25 & 3 & TIMEOUT & \underline{1.07} & 1.071 \\
    37 & 1000 & 50 & 25 & 4 & TIMEOUT & 3.575 & \underline{3.573} \\
    38 & 1000 & 50 & 25 & 5 & TIMEOUT & \underline{14.424} & 16.327 \\
    39 & 1000 & 50 & 25 & 6 & TIMEOUT & \underline{56.359} & 59.671 \\
    40 & 1000 & 250 & 125 & 6 & TIMEOUT & \underline{0.569} & 0.577 \\
    41 & 1000 & 100 & 50 & 8 & TIMEOUT & 21.447 & \underline{21.414} \\
    \hline\noalign{\smallskip}
  \end{tabular}
\end{table}

\textbf{Test problem description.} All the ILP
problems with binary variables from a given experiment have
artificially generated quasi-block structures. All the blocks from a
single problem have the same number of variables, and also the same
number of variables in separators between them. This is required in
order to evaluate the impact of the PA on the time to solve the
problem by increasing the number of variables.

The test problems were generated by specifying the number of
variables, the number of constraints and the size of the separators
between blocks. The number and the dimensions of the blocks were
calculated by using the number of variables and constraints. The
objective function and constraint matrix coefficients, and the
right-hand sides for each of the block were generated by using a
pseudorandom-number generator.

Each test problem was solved by using three algorithms,
a) the basic MILP SYMPHONY solver with the $OsiSym$ interface,
b) the LEA in combination with SYMPHONY,
c) the LEA in combination with SYMPHONY and with PA (warm restarts).
In all the cases SYMPHONY used preprocessing.

The computational experiments show that LEA combined with SYMPHONY for
solving quasi-block problems with small separators outperforms the
stand alone SYMPHONY solver (see table
\ref{benchmarking_table}). Additionally, by increasing the size of the
separators in the problems for the same number of variables and block
sizes LEA becomes less efficient due to the increased number of
iteration for solving the block subproblems. LEA's efficiency is
improved by using warm restarts. The ILP problems corresponding to the
same block for different values of the separator variables differ only
in the right-hand side. These problems can be solved partially by
using warm restarts and information obtained from other problems, and
this was expected to increase the LEA performance. However, the
results show a inconsistent behavior.  For most problems, the warm
restarts don't make a difference. For some problems, they improved the
solution time (see problems 7, 17, 24), while for others they did not
(see problems 5, 15, 40).

Concluding, there is not a definite answer for the block separator
size that results to LEA's lower performance (with respect to time to
solution) compared to SYMPHONY. The performance depends on the problem
structure; as the problem size increases LEA becomes more efficient.

\section{Conclusion}

\textit{The main result of this paper is} to determine the real computational
capabilities of block elimination algorithms combined with SYMPHONY
solver. Analysis of the results show that for sufficiently large
number of blocks and small enough size of separators between the
blocks for staircase integer linear programming problem the local
elimination algorithms in combination with a solver for solving
subproblems in blocks allow to solve such problems much faster than
used solver itself for solving the whole problem.

It seems promising to continue this line of research by studying
capabilities of postoptimal analysis distributed by other modern
solvers. It is also of interest to research computational capabilities
of local algorithms for solving sparse problems of integer linear
programming from real applications.

\end{document}